\documentclass[conference]{IEEEtran}
\IEEEoverridecommandlockouts
\usepackage{cite}
\usepackage{amsmath,amssymb,amsfonts}
\usepackage{flushend}
\usepackage{algorithmic}
\usepackage{graphicx}
\usepackage{textcomp}
\usepackage{xcolor}
\usepackage{subfig}
\usepackage{svg}
\usepackage{caption}
\def\BibTeX{{\rm B\kern-.05em{\sc i\kern-.025em b}\kern-.08em
    T\kern-.1667em\lower.7ex\hbox{E}\kern-.125emX}}
\begin{document}

\title{Automated Spectrum Sensing and Analysis Framework}

\author{
\IEEEauthorblockN{Rahil Gandotra\IEEEauthorrefmark{1}, 
Ruoyu Sun\IEEEauthorrefmark{1}, 
Mark J. Poletti\IEEEauthorrefmark{1},
Jiayu Mao\IEEEauthorrefmark{2} and
Hao Guo\IEEEauthorrefmark{1}}

\IEEEauthorblockA{\IEEEauthorrefmark{1}\textit{Wireless Technology, CableLabs}, Louisville, Colorado, USA\\
email: \{r.gandotra, r.sun, m.poletti, h.guo\}@cablelabs.com}

\IEEEauthorblockA{\IEEEauthorrefmark{2}\textit{Department of Electrical and Computer Engineering, The Ohio State University}, Columbus, Ohio, USA\\
email: mao.518@osu.edu}
\thanks{This paper is under review at an IEEE conference.}
}

\maketitle

\begin{abstract}
Spectrum sensing and analysis is crucial for a variety of reasons, including regulatory compliance, interference detection and mitigation, and spectrum resource planning and optimization. Effective, real-time spectrum analysis remains a challenge, stemming from the need to analyse an increasingly complex and dynamic environment with limited resources. The vast amount of data generated from sensing the spectrum at multiple sites requires sophisticated data analysis and processing techniques, which can be technically demanding and expensive. This paper presents a novel, holistic framework developed and deployed at multiple locations across the USA for spectrum analysis and describes the different parts of the end-to-end pipeline. The details of each of the modules of the pipeline—data collection and pre-processing at remote locations, transfer to a centralized location, post-processing analysis, visualization, and long-term storage—are reported. The motivation behind this work is to develop a robust spectrum analysis framework that can help gain greater insights into the spectrum usage across the country and augment additional use cases such as dynamic spectrum sharing.
\end{abstract}

\section{Introduction}
The radio frequency (RF) spectrum refers to the range of electromagnetic frequencies, typically ranging from about 3 kHz to 300 GHz, that are primarily used for wireless communication. Virtually all wireless technologies – mobile communication (2G, 3G, 4G, 5G, and 6G), wireless data (Wi-Fi, Bluetooth), broadcasting (AM/FM radio, television), navigation (GNSS), specialized uses (Radar, satellite communication, air traffic control, emergency services communications) – are enabled through RF spectrum. However, RF spectrum is a finite natural resource, making it one of the most sought-after resources in the world. Since radio waves can interfere with each other if they use the same frequencies in the same area at the same time, their use is carefully managed and regulated. For example, in the United States, the responsibility of RF spectrum allocation is divided between two key government agencies – the Federal Communications Commission (FCC) and the National Telecommunications and Information Administration (NTIA). The FCC, an independent regulatory agency, allocates and regulates  RF spectrum for non-federal use (i.e., state and local governments and commercial uses), while the NTIA, an agency under the Department of Commerce, allocates  RF spectrum for federal use (e.g., use by the Department of Transportation, and numerous additional federal agencies) \cite{b1}.

An essential component of effective spectrum management is spectrum analysis, the process of observing, measuring, and analyzing RF transmissions. Spectrum analysis is crucial for several key reasons: detecting and mitigating interference, ensuring regulatory compliance, protecting critical infrastructure (like GPS, satellite, and public safety networks), understanding spectrum utilization, and supporting new technologies. Consequently, depending on the specific use case, there are different types of spectrum analysis activities: compliance monitoring, interference hunting and mitigation, spectrum utilization analysis, signal intelligence, and quality assurance. The focus of this paper is on spectrum utilization analysis – measuring how much of the spectrum is actually being used. The analysis of spectrum utilization provides valuable insights that help optimize policy and spectrum management decisions. Understanding spectrum utilization also aids in determining possibilities for spectrum sharing, allowing different services to coexist in the same frequency bands. Spectrum utilization data can also inform regulators about the status of current spectrum bands and enable them to evaluate future sharing policies. Moreover, efficient dynamic spectrum sharing (DSS) also requires the utilization of data for systems to opportunistically use unused frequencies without causing interference to primary users. Thus, spectrum utilization analysis provides the foundation for making spectrum management more agile, analytical, and efficient.

When designing a spectrum management system, the three fundamental dimensions to start with are frequency, time, and geographic area, and any effective spectrum management system should take into account these three factors appropriately to provide useful information. However, analyzing the spectrum in such a complex wireless environment over a wide frequency range, in real time and across many geographic areas remains a demanding challenge. One-time, single-site surveys do not provide sufficient data to be of value. Additionally, high-quality spectrum analysis equipment is expensive, and maintaining a comprehensive analysis network requires significant ongoing investment in equipment, personnel, and facilities. Further adding to the complexity is the enormous amount of data generated that needs to be stored and analyzed. Thus, there are many technical and operational challenges that can significantly impact spectrum analysis.

In this paper, we present a novel automated spectrum analysis framework developed and deployed in multiple locations in the USA, extending the previous work in [2,3], and describe the different parts of the end-to-end pipeline. The details of each of the modules of the pipeline—data collection and pre-processing at remote locations, transfer to a centralized location, post-processing analysis, visualization, and long-term storage—are discussed in Section II. In Section III, we report the results of the analysis gathered from the collected spectrum data. Section IV reviews the recent advances in spectrum analysis. Section V concludes the paper and provides directions for future work. The motivation behind this is to develop a robust spectrum analysis framework that can help gain greater insights into the spectrum usage across the country and augment additional use cases such as DSS.

\section{Automated spectrum analysis framework}
RF spectrum analysis is a multi-step process involving interconnected components, both hardware and software, that work together to create a comprehensive analysis system. Fig. 1 shows the major components and steps involved in the spectrum utilization analysis pipeline. The first step in the pipeline is signal detection by an antenna and a spectrum scanner. In case of spectrum being sensed at multiple geographic locations, the next step is to transfer the data from each remote location to a central location. Once the data has been collected at the central location, the next step is to perform analysis to extract the spectrum utilization patterns for sharing studies. The next step is to visualize the analyzed data to simplify the complex information and enhance data understanding. Also, all the collected data needs to be organized and stored for historical analysis. In addition to all these steps, the whole pipeline needs to be managed and maintained to minimize downtime and preserve data accuracy.

\begin{figure}[htbp]
\centerline{\fbox{\includegraphics[width=0.95\columnwidth]{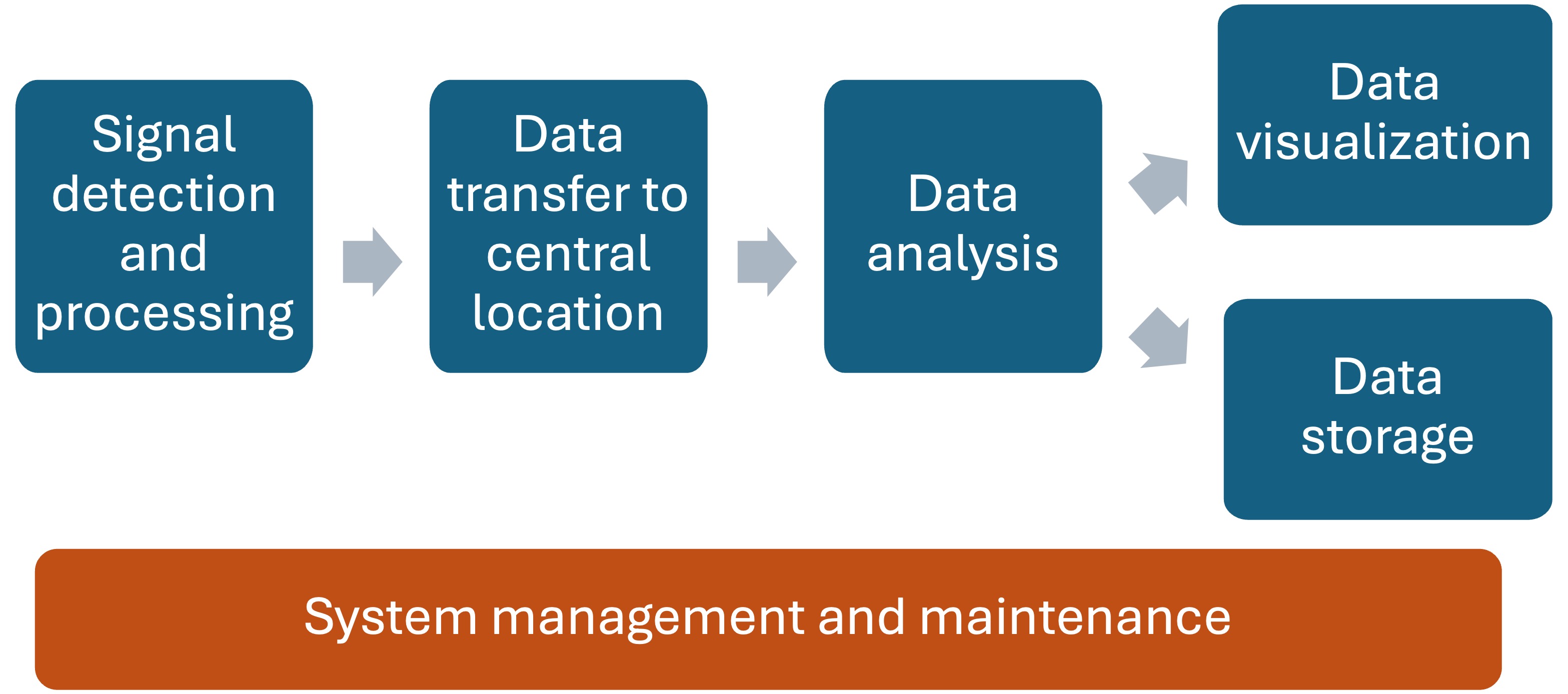}}}
\caption{Spectrum analysis framework}
\label{fig}
\end{figure}

Manual RF spectrum analysis, while historically common and still useful for specific, localized tasks, faces significant challenges in today's increasingly complex and congested wireless environment. With the vast amount of data generated and required to analyze, manual processes become practically impossible. To perform effective real-time spectrum analysis across multiple locations, automation becomes not just beneficial but essential. Automation streamlines repetitive tasks thereby significantly reducing the time and effort required for these operations, allowing greater focus on more strategic initiatives. Manual operations are prone to human error and automation minimizes these risks by ensuring consistent and accurate execution of tasks. Additionally, automation enables spectrum analysis to scale more easily to meet growing demands. Automated provisioning and configuration allow for greater agility, making it simple to add or remove new locations, enabling, as a result, the spectrum analysis framework to adapt quickly to changing business requirements and technological advancements.

Fig. 2 depicts the automated spectrum utilization analysis system comprising of multiple remote locations and a central location (CableLabs, Louisville, Colorado, USA). The system comprises of different modules deployed both at the remote locations as well as the central location, that work together in an automated way to enable spectrum analysis at scale.

\begin{figure}[htbp]
\centerline{\fbox{\includegraphics[width=0.95\columnwidth]{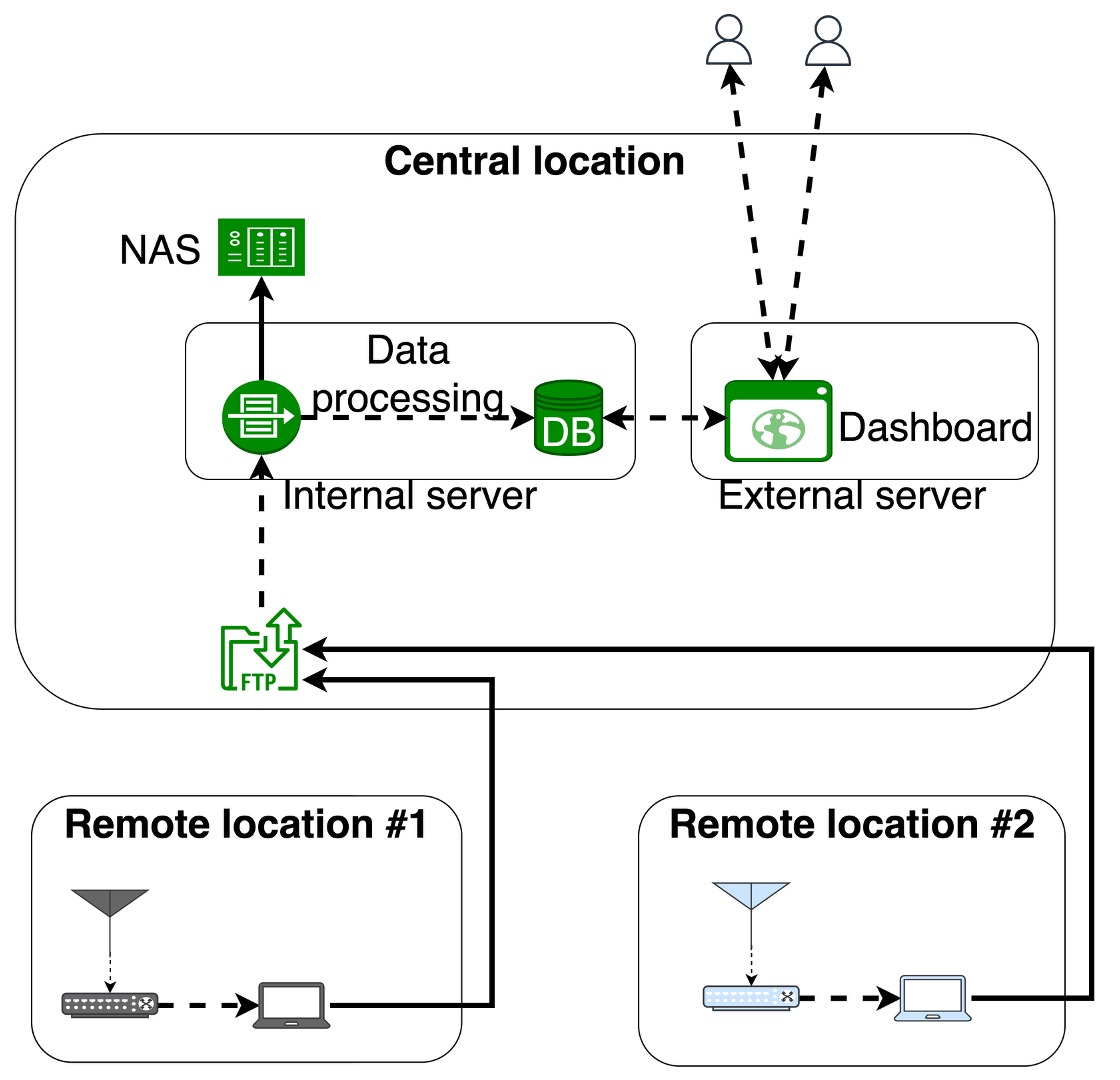}}}
\caption{Automated spectrum utilization analysis across multiple locations}
\label{fig}
\end{figure}

\subsection{Signal Detection and Recording in Remote Locations}
The components for signal detection and processing were described in detail in \cite{b2}. In brief, an antenna is installed at the remote location of interest, which is connected to a spectrum scanner, Signal Hound BB60C, using a coaxial cable. In sites with long distances between antenna and spectrum scanner, a low-noise amplifier (LNA) can be added to compensate for cable loss. Additionally, a lightning protector is also installed to safeguard the equipment. Based on the cable length between antenna and spectrum scanner, the system is calibrated to determine and compensate for the cable loss and non-ideal nonlinearity over frequency.

A local computer, either a laptop or a single-board computer, records and compresses the collected data. To enable 24/7 data collection, a power threshold is applied for each location, which detects and stores signal data only above the specified threshold. A Python script running on the computer uses the APIs provided by the spectrum scanner to configure the required settings for RF scanning – frequency range, resolution bandwidth, sweep time, and power threshold. The data is stored in the Parquet format, which is an open-source columnar storage file format optimized for efficient data storage and retrieval, particularly for large datasets. Each file contains the UNIX timestamp, frequency, power level, and other system parameters. The system parameters used for data collection for the past hour are stored in a JSON file listing the frequency range, resolution bandwidth, sweep time, which is a measure of the minimum acquisition time, geographic coordinates of the remote location, antenna type deployed at the location, and LNA type used. The JSON file can be used during data analysis to derive functional metrics. Parquet files are partitioned by minute, while JSON files are recorded hourly.

\subsection{Data Transfer to Central Location}
In order to further analyze the spectrum data, it needs to be transferred to the central location from each remote location. Since the amount of raw data generated at each site is huge, ranging from 30 GB to 200 GB per day, depending on activity within the sensed frequency range at that location, it is inefficient to send the raw data as is from all the remote locations to the central location. A Python script running on the computer at each remote location first aggregates the raw spectrum data and the associated system parameters used for collection for the past day, compresses them into an archive, and then transfers the compressed archive to an FTP server located at the central location.

To compress the raw spectrum utilization data, a lossless compression algorithm needs to be used to preserve all original data which allows for perfect restoration. There exist multiple lossless compression methods such as ZIP, GZIP, BZIP2, and 7-ZIP, which use different compression algorithms. 7ZIP, which uses the Lempel-Ziv-Markov chain Algorithm (LZMA), generally provides the best compression ratio \cite{b4}. In data compression, there’s a fundamental trade-off between the time it takes to compress the data and the resulting compression ratio. Since for our use case, achieving higher compression ratio takes precedence over faster compression time, 7-ZIP was used with the best compression level (9) which achieved a compression ratio around 5:1, i.e., the raw data was compressed to about 20\% of its original size. This allows for faster transfer times from the remote locations to the FTP server, while consuming a lower amount of bandwidth.

Once the compressed archive is successfully transferred to the FTP server on the central location, the Python script deletes both the raw data and the compressed archive in order to not fill up the storage. The Python script is scheduled to trigger every day using the job scheduler available on the computer, such as the Windows Task Scheduler.

\subsection{Data Analysis}
Once the compressed archive has been received on the FTP server, a Python script running in one of the internal servers in the central location moves the compressed archive from the FTP server. The Python script is triggered to run every day for each remote location using a job scheduler. After the transfer, the Python script next uncompresses the 7-ZIP archive to extract the raw spectrum utilization data. The raw spectrum data is then processed by invoking the MATLAB Engine API from the Python script which is enabled using MATLAB as the computational engine. Depending on the remote location, the Python script provides additional inputs such as the time zone of the remote location, the power thresholds to be used, and the channel bandwidth.

As described in \cite{b2}, to effectively analyze the data, two essential metrics were introduced: channel occupancy and airtime utilization (AU), which are presented in Section III. These results are saved in a CSV file and then stored by the Python script into a database (MYSQL) in a structured format making it easy to locate, retrieve, and manage the analyzed data.

%To compute channel occupancy, the measured spectrum data is divided into channels with channel bandwidth (e.g., 5 MHz), and a channel is considered to be occupied during any one-second period if it records one or more data points above the power threshold. For each location, the system considers two power thresholds – one corresponding to -72 dBm/20MHz which matches the power detection (PD) threshold use in the IEEE 802.11 Clear Channel Assessment (CCA) technique, and the other corresponding to -89 dBm/MHz which matches the detection threshold used by the CBRS Environmental Sensing Capability (ESC). Airtime utilization is calculated by measuring the percentage of time a channel is occupied relative to the total available time (e.g., airtime utilization per hour is the ratio of the total occupied seconds within 3600 seconds). The final airtime utilization results are saved in a CSV file and then stored by the Python script into a database (MYSQL) in a structured format making it easy to locate, retrieve, and manage the analyzed data.

\subsection{Data Visualization}
The next step in the pipeline is to visualize the analyzed airtime utilization data effectively to convey complex information to diverse audiences. By converting data into visual representations, data visualization helps users quickly spot trends, patterns, and outliers that might be hidden in the data. The goal is to transform intricate datasets into easily digestible formats, making information accessible to a wider audience, enabling faster and more informed decision-making.

There are primarily two methods to develop and share the data visualization module – desktop-based and web-based. While desktop-based visualization is simplified to develop and set up, it suffers from certain limitations due to its platform dependence. Web-based visualization can be accessed from any device with an internet connection and a web browser, eliminating the need for specific software installations or operating system compatibility. They can also provide centralized access control, streamlining user management and permissions. Additionally, a web-based module typically employs a single, central version, ensuring that all users are accessing the latest updates and data. The interactivity of web-based tools, leveraging the capabilities of HTML5 and JavaScript to create highly interactive and engaging visualizations, make them more memorable and impactful. Therefore, a web-based data visualization module was developed within our spectrum utilization analysis framework. Since the dashboard needs to be publicly accessible over the internet to allow all stakeholders from different remote locations to be able to see the visualized data, the dashboard was installed on an external-facing server, effectively creating different network zones to have a buffer between the LAN and the WAN, to protect the LAN from untrusted traffic originating from the WAN.

The web-based dashboard was developed in Python using Flask, a lightweight and flexible micro web framework \cite{b5}. Flask's minimalist design makes it suitable for both small-scale projects and rapid prototyping, while its extensibility allows it to be scaled for more complex applications. The Flask-Login package available within its framework was employed to enable authentication to the dashboard. This allows for managing user sessions, hashing, and storing passwords in a database, creating sign-up and login forms for users to create accounts and log in, and defining protected pages for logged-in users only.

After logging in, the user is directed to a landing page which depicts all the remote locations on a geographic map, from which the location of interest can be selected. Mapbox was used to create maps, leveraging their suite of APIs, Software Development Kits (SDKs), and HTML packages to create custom, high-performance location-based features \cite{b6}. Fig. 3 shows the landing page after a user logs in. The dashboard additionally provides the user with the ability to filter on the dates,  frequency range, and power threshold of interest. These user inputs are used to filter and obtain the required data from the database. To visualize the spectrum utilization data for any remote location, Plotly was used, which is an open-source Python graphing library used for creating interactive visualizations \cite{b7}. Plotly provides a high-level, declarative interface for generating a wide variety of chart types. To facilitate a more comprehensive understanding of the spectrum utilization data, multiple different types of graphs were developed to highlight specific insights and present different facets of the data.

\begin{figure}[htbp]
\centerline{\fbox{\includegraphics[width=0.95\columnwidth]{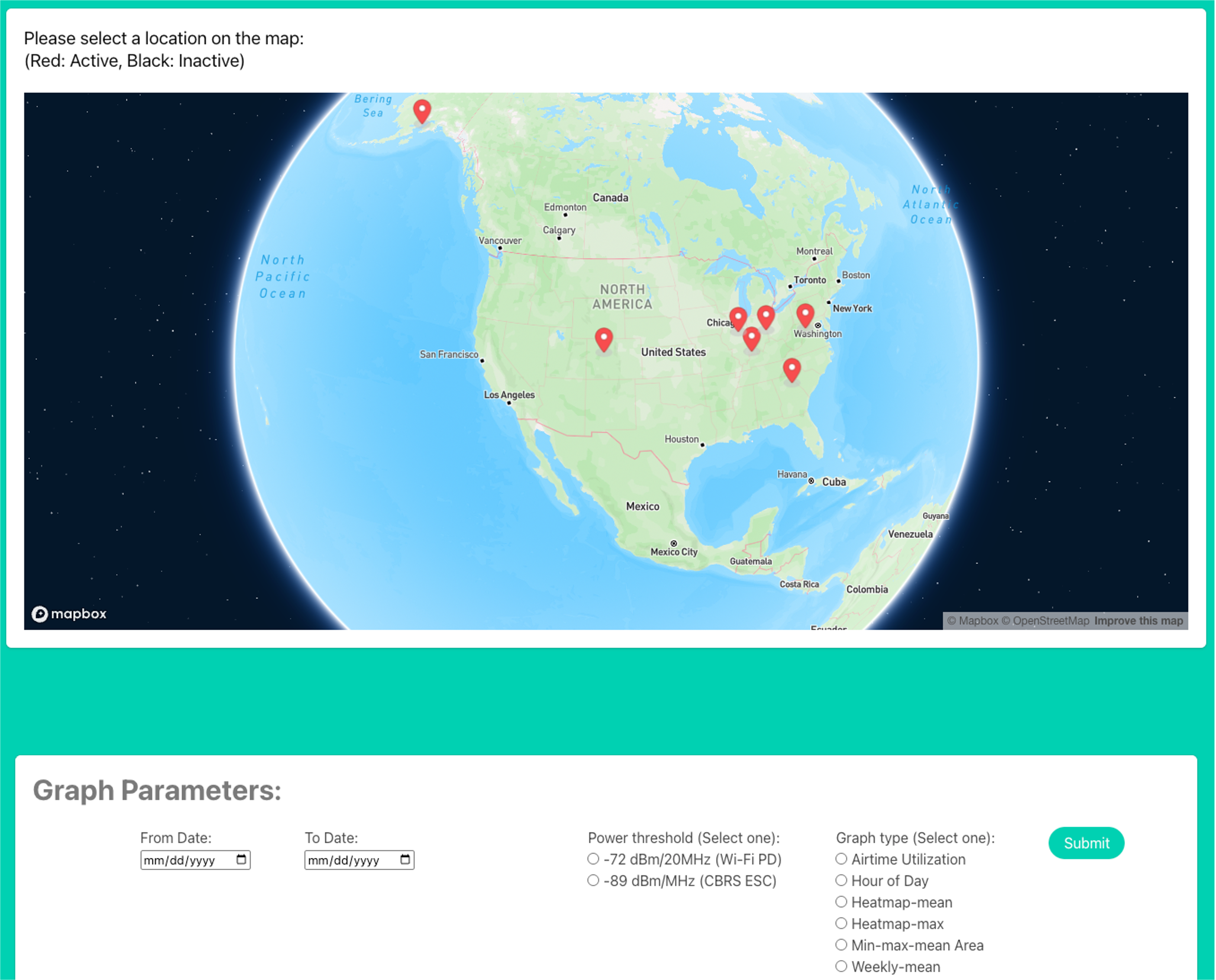}}}
\caption{Landing page of the dashboard after a user logs in}
\label{fig}
\end{figure}

The graph types currently supported are – airtime utilization, hour of day, heatmap-mean, heatmap-max, min-max-mean area, and weekly-mean as presented in \cite{b3}. Some example airtime utilization data are illustrated in Section III. 

%Airtime utilization is a graph depicting the airtime utilization computed for a location for the specified time and frequency range. Both two-dimensional (2D) and 3-dimensional (3D) interactive graphs are presented which allows the user to zoom, pan and hover over data points for details. The hour of day graph shows the average airtime utilization for every hour of the day (0-23) for the specified time and frequency range in 2D and 3D, which is ideal for understanding trends and changes in data over the 24-hour period. For the two heatmap graph types, the continuous values of airtime utilization are binned into four discrete intervals – highly utilized (50\%-100\%], moderatelydefense utilized (20\%-50\%], underutilized (0\%-20\%], and not utilized [0\%] – for each hour of day. The heatmap-mean graph shows the heatmap of the mean of the airtime utilization values, while the heatmap-max shows the heatmap of the max of the airtime utilizations. Heatmaps, using their use of different bin colors, allow to see data with low granularity providing an overall birds-eye view. The min-max-mean area graph shows the minimum, maximum, and mean airtime utilization values for the specified time and frequency range. The weekly-mean graph shows the average airtime utilization for the week of the year (1-52) for the specified time and frequency range in 2D and 3D, which is ideal for understanding seasonal trends and changes in data across months.

\subsection{Data Storage}
Long-term storage of spectrum data is crucial for several strategic and operational reasons that extend far beyond immediate analysis needs. Long-term data enables identification of spectrum usage patterns that emerge over months or years. Gradual shifts in technology adoption can also become visible through historical data. This information is also invaluable for predicting future spectrum demand and planning allocation strategies. Regulators need comprehensive data spanning multiple years to make informed decisions about spectrum reallocation, auction planning, and policy changes. Long-term storage provides the evidence base for major spectrum policy decisions that affect entire industries and billions of dollars in infrastructure investments. Furthermore, researchers studying radio propagation, spectrum efficiency, or wireless technology development rely on long-term datasets to validate theories and develop novel approaches. Historical spectrum data becomes a valuable scientific resource that supports innovation and understanding.

In the developed automated pipeline, the last step after the data has been analyzed and stored in the database is to move the compressed raw spectrum data to a network-attached storage (NAS) server. The Python script deletes the uncompressed raw data on the internal server, to prevent its storage from exceeding its capacity limit, and then transfers the compressed raw data into a folder on the NAS server for the specific remote location. In this way, the historical data for all remote locations is available to support any analysis requirements in the future. To give an example, the total size of the compressed data for one month for seven remote locations is approximately 3.5TB, therefore substantiating the need to compress data for transfer and storage. Since collecting data over extended periods for multiple locations can lead to storage exhaustion, the NAS server used is extendable to allow for adding more storage disks to increase its storage capacity.

\subsection{System Management and Maintenance}
System management and maintenance of spectrum analysis systems require comprehensive, ongoing work across all components to ensure reliable operation, accurate measurements, and effective performance. Since there are multiple modules within the developed analysis framework, managing each module is critical to maintain the system’s availability and data integrity. To achieve this, an accompanying operations system was developed with the capability to observe the ongoing processes and inform the system administrators in case of any failure. This is essential to maintain and scale to multiple locations, so that failures can be identified and rectified immediately to minimize system downtime.

The operation systems developed consists of a web-based dashboard that can be used by the system administrators to check the current status of the system including all the remote locations. Several associated scripts are used in the backend to monitor the current status and populate it correspondingly in the dashboard. Also, a notifications system is also developed to apprise the administrators in near-real time. The operations system monitors three distinct types of potential failures for all remote locations – (i) the computer has become unreachable, or (ii) data collection has stopped on the remote computer, or (iii) the compressed data archive was not successfully transferred to the central location. The local computer can become unreachable due to loss in network connectivity or power at the remote location. Data collection can stop on the remote computer due to any local error in the transfer between the spectrum scanner and the computer or if the computer was restarted due to any reason, such as an OS update. Any failure in the FTP server at the central location or in the Python script on the remote computer triggered to transfer the compressed archive can cause it not to be successfully transferred.

To identify the first failure, the computer has become unreachable, a Python script running on the computer of each of the remote locations is triggered at set intervals to transfer a small file to the FTP server at the central location. Since the computer at each remote location is typically installed in the LAN behind a firewall, it is unfeasible to test connectivity to the computer from outside. Hence, the mechanism developed, like a heartbeat, is for the computer to periodically generate a signal indicating normal operation. Another Python script running on the internal server at the central location is also triggered at set intervals to check if the file is received at the FTP server within a certain time frame for each remote location. In case the file is not received within that time frame, the corresponding status is updated in the operations dashboard. Additionally, a notification is triggered using Slack to inform the administrators. The Slack Python SDK is used to invoke APIs provided by Slack to send messages to a dedicated Slack channel.

To identify the second failure, data collection has stopped on the remote computer, a Python script running on the computer of each remote location is triggered at set intervals to check if the data collection Python process is running. If the process is running, the Python script transfers a small file to the FTP server at the central location. Another Pythons script running on the internal server at the central location is also triggered at set intervals to check if the file is received at the FTP server within a certain time frame for each remote location. In case the file is not received within that time frame, the corresponding status is updated in the operations dashboard. Additionally, a notification is triggered in Slack to inform the administrators, based on which corrective action can be taken.

To identify the third failure, compressed data archive not successfully transferred to the central location, a Python script running on the internal server at the central location is triggered at set intervals to connect to the FTP server to check the file modified time of the latest received file, for each remote location. If the file modified time is found to be greater than a day, that indicates a failure in the transfer of the compressed archive, and the corresponding status is updated in the operations dashboard, and a notification is triggered in Slack informing the administrators to investigate the issue.

Using all of these proactive automated failure monitoring and notifications methods allow the system to be fully available. Minimal disruption in the end-to-end automated pipeline ensures there is no gap in the data collected across the time dimension. Additionally, to operate the system at scale to manage multiple remote locations, the developed operations system shoulders much of the maintenance load allowing the administrators to focus on critical tasks.

\section{Results Analysis}
\begin{figure}[ht]
    \centering
    \includegraphics[width=1\linewidth]{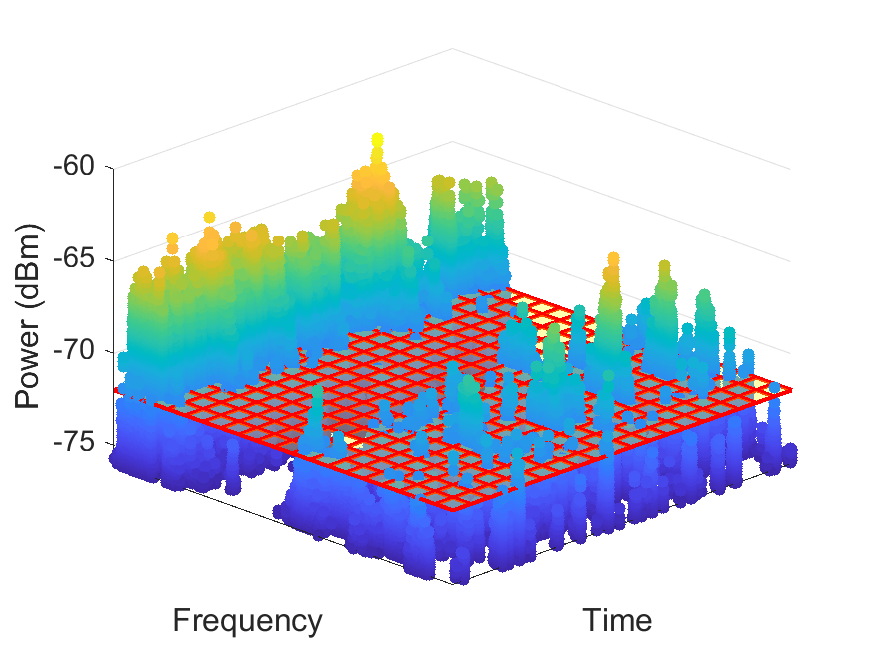}
    \caption{3D raw data}
    \label{fig:raw3D}
\end{figure}

As briefly outlined in Section II.C and comprehensively detailed in \cite{b2} and \cite{b3}, airtime utilization is employed to characterize spectrum activity and assess the availability of channels for potential future sharing. The monitored frequency span is divided into 5-MHz-wide channels, and AU is calculated for each channel over one-hour time intervals. AU represents the percentage of time (or sweeps) during which a given channel is occupied, relative to the total observation time (or number of sweeps), with values ranging from 0\% to 100\%. Channel occupancy is determined in a binary manner (either occupied or unoccupied) depending on whether at least one frequency point inside a 5-MHz channel exceeds a defined power threshold. For this study, the -72 dBm/20 MHz power threshold is applied to evaluate spectrum utilization, consistent with the Wi-Fi listen-before-talk (LBT) criterion. A segment of example raw measurement data is illustrated in Fig.~\ref{fig:raw3D}, where the -72 dBm/20MHz power threshold (red plane) is depicted as a red plane. The rectangular grid (not to scale) overlaid on this red plane represents the frequency and time bins used in the computation of channel occupancy.

\begin{figure*}
    \centering
    \includegraphics[width=1\linewidth]{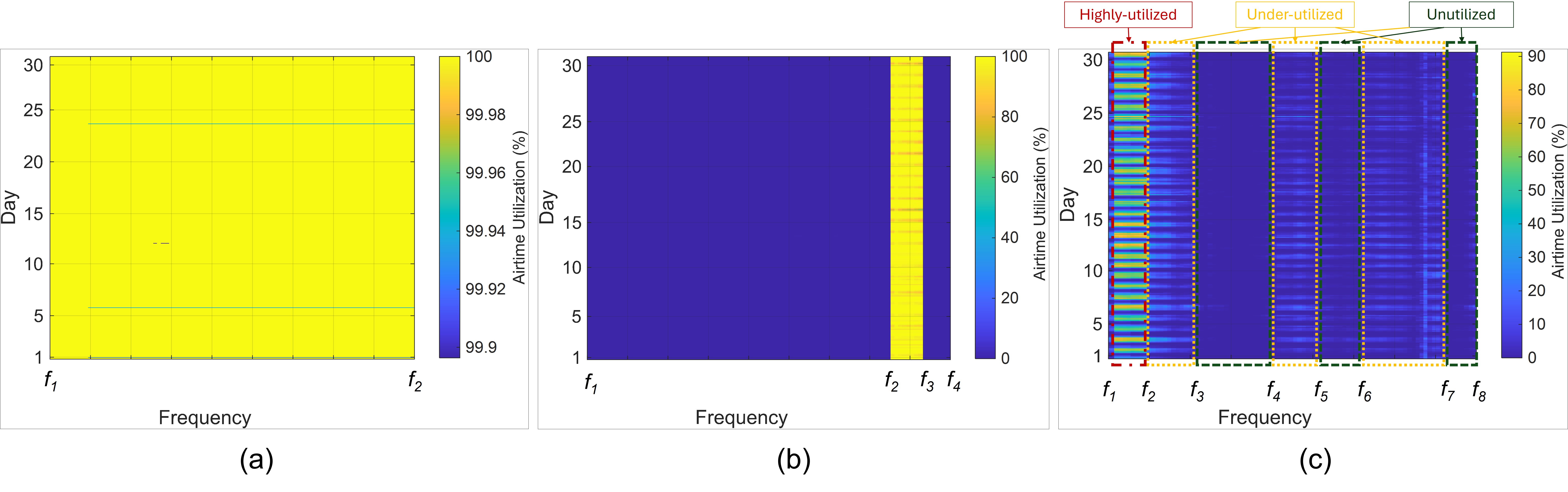}
    \caption{Example airtime utilization results collected in June 2025 over 450-MHz in mid-band at (a) Anchorage, AK; (b) Louisville, KY; (c) Louisville, CO}
    \label{fig:AirtimeUtilization}
\end{figure*}

Fig.~\ref{fig:AirtimeUtilization}(a) shows the AU in downtown Anchorage, AK, on the rooftop of a GCI office building, spanning a 450-MHz frequency range in mid-band with 5-MHz granularity, and covering the period of June 1-30, 2025, with one-hour temporal granularity. The color scale in Fig.~\ref{fig:AirtimeUtilization} indicates airtime utilization from 0\% to 100\%. The data collected on the NCTA building in downtown Washington DC are very similar to Fig.~\ref{fig:AirtimeUtilization}(a). The monitored band is heavily used in these two coastal locations. The AU is almost 100\% all the time.

Fig.~\ref{fig:AirtimeUtilization}(b) demonstrates AU measured on the University of Louisville, Kentucky campus, which is remarkably similar to the data observed at Purdue University in West Lafayette, Indiana and The Ohio State University in Columbus, Ohio. At all three sites, no signal was detected  \textit{$f_1$}-\textit{$f_2$} and \textit{$f_3$}-\textit{$f_4$} throughout the observation period. In contrast, a 40 MHz segment from \textit{$f_2$}-\textit{$f_3$} exhibited consistently high utilization, with almost 100\% airtime utilization across all time.

These findings suggest a localized and consistent pattern of high utilization in a narrow portion of the \textit{$f_2$}-\textit{$f_3$} band, while the remainder of the mid-band spectrum remains largely unused at these in-land locations. This underutilization presents an opportunity for spectrum sharing to support emerging applications including commercial cellular networks, smart agriculture, intelligent transportation, and other wireless applications.

The AU data collected from the rooftop of the CableLabs headquarters in Louisville, Colorado are illustrated in Fig.~\ref{fig:AirtimeUtilization}(c). The frequency range from \textit{$f_1$}-\textit{$f_2$} is the most heavily utilized portion of the spectrum with AU sometimes exceeding 90\%. In contrast, the \textit{$f_2$}-\textit{$f_3$}, \textit{$f_4$}-\textit{$f_5$}, and \textit{$f_6$}-\textit{$f_7$} bands are under-utilized, with the max AU below 40\% and median AU under 5\%. Furthermore, the \textit{$f_3$}-\textit{$f_4$}, \textit{$f_5$}-\textit{$f_6$} and \textit{$f_7$}-\textit{$f_8$} bands are almost unutilized, with a median AU below 1\%. These unutilized and underutilized segments present opportunities for potential spectrum sharing.

The heavily utilized segments, such as the \textit{$f_1$} to \textit{$f_2$} band, show a strong dynamic temporal pattern: higher activity during daytime hours and significantly reduced usage in the evening (a detailed analysis was presented in \cite{b2} and \cite{b3}). This diurnal variation suggests the potential for dynamic spectrum sharing, allowing the spectrum to be repurposed during periods of low utilization to support emerging applications such as smart agriculture or power-line inspection, and other intermittent or delay-tolerant services.

The automated spectrum analysis framework developed enabled the findings presented in this section. Automation enables the ability to scale, adapt and centralize data collection and analysis, resulting in a more comprehensive view of the spectrum environment.

\section{Related Work}
Recent advances in automated spectrum monitoring have prioritized deployable, end-to-end systems that integrate wideband sensing, local digital signal processing, scalable data transport, cloud analytics, and interactive visualization. Electrosense, a global crowd-sourced monitoring network built on low-cost SDR sensors, autonomously performs FFT-based feature extraction and stream measurements to a cloud backend for real-time signal detection, anomaly analysis, long-term storage, and visualization \cite{b8}. Its successor, Electrosense+, extends the pipeline with on-device DSP, encrypted remote-access interfaces, and user-controlled demodulation services, providing an operational blueprint for fully automated spectrum-monitoring infrastructures that are practical at scale \cite{b9}.

Complementary large-scale systems have been demonstrated in the DARPA Spectrum Collaboration Challenge (SC2), where autonomous AI-driven radios implemented complete sensing-to-decision loops including real-time spectrum detection, classification, and interference mitigation—showcasing some of the capabilities enabled by automated monitoring and coordination in congested bands \cite{b10}. Additionally, the Spectrum Observatory provided one of the earliest long-duration, multi-site measurement platforms with automated scanning schedules, data ingestion, and visualization pipelines, influencing subsequent regulatory and research deployments \cite{b11}. Together, these systems show that successful automated spectrum monitoring requires robust integration of distributed sensing hardware, adaptive DSP, secure data streaming, cloud-native analytics, and user-facing tools for situational awareness. The framework presented in this paper extends the prior work by incorporating multi-day archive compression and transfer workflows, structured database-backed analytics aligned to spectrum-utilization metrics, and an operations system for maintaining large deployments, enabling continuous multi-site sensing with multi-month statistical characterization, therefore making it closer to a production-grade, fault-tolerant distributed sensing platform.

\section{Conclusion}
The RF spectrum is the fundamental basis for our interconnected wireless world, enabling countless technologies that we rely on daily. However, it is a finite resource that must be carefully managed; different portions are allocated for specific purposes – broadcasting, mobile communications, military use, scientific research, and emergency services communications. RF spectrum analysis is crucial for efficient spectrum management due to the limited and congested nature of RF spectrum. Spectrum analysis is the systematic process of observing, measuring, and analyzing RF transmissions using specialized equipment and techniques to detect, identify, and characterize radio signals in a given area or frequency range. Performing effective spectrum analysis across the three dimensions of time, frequency and geographic area faces numerous technical and operational challenges that can significantly impact its success. This paper presents an end-to-end automated spectrum utilization framework, that addresses these difficulties by using various software components to transform spectrum analysis from a reactive, labor-intensive, and often incomplete process into a proactive, continuous, and highly efficient operation. The framework is developed to be modular, allowing for independent evolution, and making the entire system open to be extendable to additional use cases. Based on new requirements and demands, individual modules can be replaced or enhanced to support other frequency ranges, data granularity, and data analysis. Automation allows the system to scale to multiple remote locations, with minimal manual work needed to onboard new locations. The associated operations system developed - monitoring each module of the pipeline, detecting failures, and notifying the administrators - ensures the entire framework faces minimum downtime, resulting in minimal disruption to the spectrum data collection process. Spectrum utilization results for various locations across the USA are also presented highlighting the framework’s capability in identifying long-term spectrum usage. The automated spectrum analysis framework developed enables performing various critical tasks related to spectrum management including identifying underutilized spectrum, revealing geographic insights, informing policy and regulation, and planning new services.

In addition to the work described in this paper, the future work for the automated spectrum analysis framework developed is to extend it to support other advanced use cases such as dynamic spectrum sharing. A key component to enable dynamic spectrum sharing is the detailed knowledge of which frequencies are actually being used at any given time and location. The analysis system developed can detect when users are actively transmitting or when they are idle. This information enables other users to opportunistically access unused spectrum. Furthermore, long-term analysis data enables machine learning algorithms to predict spectrum availability patterns, allowing dynamic systems to make more intelligent sharing decisions and even pre-position resources before spectrum becomes available. The primary motivation of this work is to transform spectrum from a static resource allocation problem into a real-time dynamic resource management solution. Spectrum analysis provides the situational awareness and control mechanisms to make this transformation possible, enabling substantially more efficient use of the RF spectrum.

%\section*{References}

\end{document}